\documentclass[conference]{IEEEtran}
\IEEEoverridecommandlockouts
\usepackage{cite}
\usepackage{amsmath,amssymb,amsfonts}
\usepackage{algorithmicx}
\usepackage{algorithm}
\usepackage{algpseudocode}  
\usepackage{amsthm}
\usepackage{subfigure}
\usepackage{graphicx}
\usepackage{textcomp}
\usepackage{amsfonts,amssymb}
\usepackage{xcolor}
\def\BibTeX{{\rm B\kern-.05em{\sc i\kern-.025em b}\kern-.08em
    T\kern-.1667em\lower.7ex\hbox{E}\kern-.125emX}}
\begin{document}
\bibliographystyle{IEEEtran}
\title{Reinforcement Learning Driven Adaptive VR Streaming with Optical Flow Based QoE}

\author{\IEEEauthorblockN{Wei Quan, Yuxuan Pan, Bin Xiang, Lin Zhang}
\IEEEauthorblockA{\textit{School of Information and Communication Engineering} \\
\textit{Beijing University of Posts and Telecommunications, Beijing, China}\\
\{louisqw, panyx, bingo6, zhanglin\}@bupt.edu.cn}
}


\maketitle

\begin{abstract}
With the merit of containing full panoramic content in one camera, Virtual Reality (VR) and 360$^{\circ}$ videos have attracted more and more attention in the field of industrial cloud manufacturing and training. Industrial Internet of Things (IoT), where many VR terminals needed to be online at the same time, can hardly guarantee VR's bandwidth requirement. However, by making use of users' quality of experience (QoE) awareness factors, including the relative moving speed and depth difference between the viewpoint and other content, bandwidth consumption can be reduced. In this paper, we propose OFB-VR (Optical Flow Based VR), an interactive method of VR streaming that can make use of VR users' QoE awareness to ease the bandwidth pressure. The Just-Noticeable Difference through Optical Flow Estimation (JND-OFE) is explored to quantify users' awareness of quality distortion in 360$^{\circ}$ videos. Accordingly, a novel 360$^{\circ}$ videos QoE metric based on PSNR and JND-OFE (PSNR-OF) is proposed. With the help of PSNR-OF, OFB-VR proposes a versatile-size tiling scheme to lessen the tiling overhead. A Reinforcement Learning(RL) method is implemented to make use of historical data to perform Adaptive BitRate(ABR). For evaluation, we take two prior VR streaming schemes, Pano and Plato, as baselines. Vast evaluations show that our system can increase the mean PSNR-OF score by 9.5-15.8\% while maintaining the same rebuffer ratio compared with Pano and Plato in a fluctuate LTE bandwidth dataset. Evaluation results show that OFB-VR is a promising prototype for actual interactive industrial VR. A prototype of OFB-VR can be found in https://github.com/buptexplorers/OFB-VR.
\end{abstract}


\section{Introduction}

\begin{figure*}[htbp]
\centerline{\includegraphics[scale=0.4]{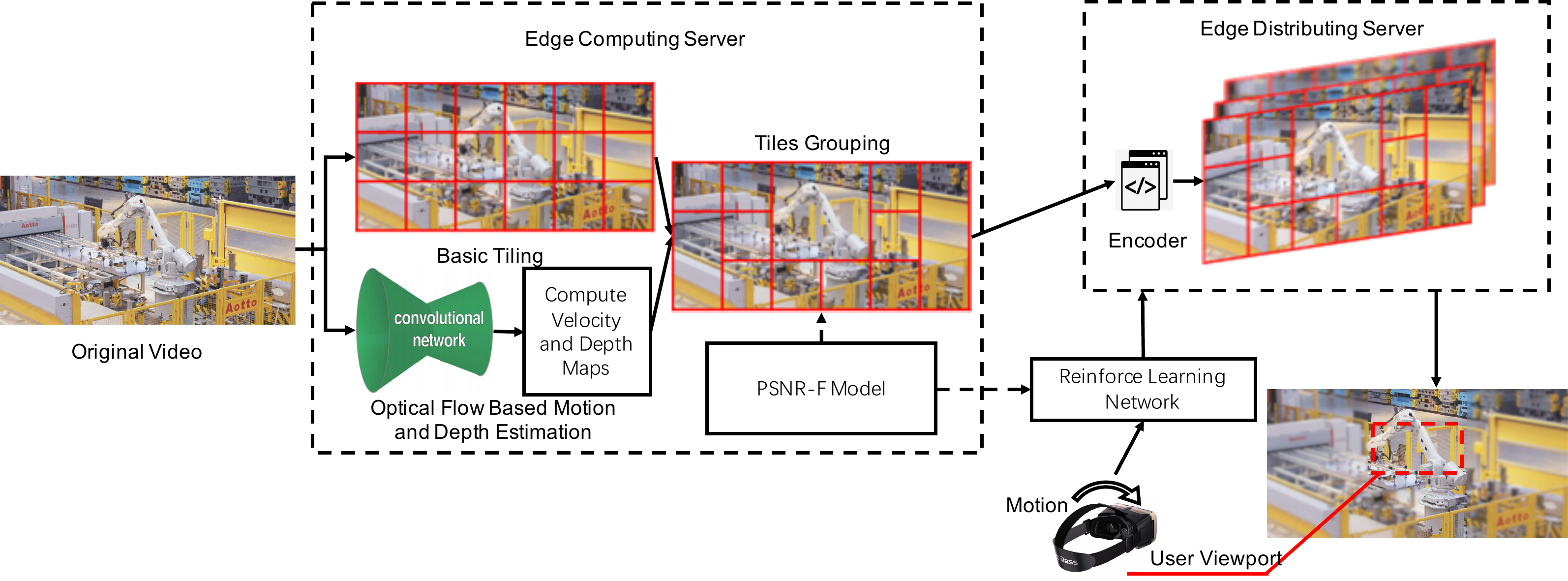}}
\caption{Flowchart of OFB-VR}
\label{Flowchart of OFB-VR}
\end{figure*}

VR technologies have grown more and more prosperous and mature recently. According to statistics, there have been 42.9 million VR users in the US(over 13\% of its population), which is predicted to grow up to 57.1 million in 2022\cite{mine:vraruser2019}. Under the overwhelming trend, content providers, including Youtube, Netflix, iQiYi, have begun to provide VR content.

As a superior form of regular videos, VR plays an irreplaceable role that the regular one cannot fit. Transmitting rich visual information by a single stream, it could reconstruct the captured scenario and make viewers feel involved and immersive. Without a doubt, entertainment is a vital and primary application scenario for VR. Except for entertaining, VR can be extremely useful in industrial areas, such as monitoring robot tasks or VR training\cite{8971967}. Cloud manufacturing and remote control can be more accurate as well. Moreover, panoramic supervising is a promising substitute for the regular one, for its omnidirectional view, ample detail, and high resolution.


Unfortunately, augmented bandwidth consumption can cripple the availability of a 360$^{\circ}$ video system. However, Tile-based streaming schemes\cite{Hosseini2017View,8486282,guan2019pano} are feasible to tackle the immense bandwidth consumption. The scheme firstly cut the original video into several equal-length segments, which are referred to as chunks, then split each chunk spatially into several minor squares, which are referred to as tiles. When streaming, the client can selectively decrease some tiles' quality to reduce bandwidth consumption. 
However, tile-based VR streaming schemes have some intrinsic limitations. Firstly, there is a trade-off between QoE robustness and QoE mean score. A fine-grained tiling scheme can reduce unviewed area efficiently to improve QoE of the viewport area, but can be more fragile to viewpoint prediction error. 
Corse-grained schemes are just the opposite. Secondly, users perceive panoramic videos in a way different from the regular one. VR users can tolerate quality distortion to some extent. In fact, the more significant the relative velocity or depth differences between the focused object and other areas are, the less likely the users notice the quality drop is.

Optical flow\cite{ilg2017flownet} is a reliable tool to estimate accurate pixel-wise motion between two continuous frames in videos. Optical flow estimation assumes that each pixel has fixed luminance and only makes small displacement, so ideally, each pixel in the previous frame can match with the one in the succeeding frame. With this method, we can explore video content related factors with precise pixel-wise motion information.

In recent years, applying RL to serve as adaptive actors in tile-based 360$^{\circ}$ videos' streaming has become a hot topic\cite{jiang2018plato}. A typical RL system consists of three parts, the action, the environment state, and the reward. Action depends on the current policy. After an action is made, actor will get a delayed reward to tell if the current action is well enough. 
During training, the algorithm optimizes policy parameters to get the cumulative reward as high as possible. Due to the merit of recalling historical decisions' reward, RL can be adopted to make optimal tile quality decisions in dramatic network variation.

This paper proposes OFB-VR, an optical flow based VR streaming system: 
\begin{itemize}
\item We use Convolution Neural Network(CNN) based optical flow estimation to extract the relative speed and velocity depth of each pixel accurately, to quantify the users' awareness of quality distortion. Then we integrate those quantified factors into PSNR to form a new QoE metric named PSNR-OF. 
\item In contrast with the fixed-size tiling scheme, We propose a versatile-size tiling scheme by grouping tiles with similar PSNR-OF efficiency together. It is proved that our scheme can help save bandwidth and increase QoE.
\item We design an asynchronous advantage actor-critic(A3C) based RL network to serve as our ABR actor. The network is designed to benefit from past information to get the highest PSNR-OF score.
\end{itemize}
This paper is organized as follows: The related work about tile-based VR streaming methods are summarized in Section II. Section III shows the system overview and detailed design about the proposed system. In Section IV, experiment setups and the parameters of our RL ABR are given, then we compare the performance of OFB-VR with two baselines, Plato and Pano. Finally, the article ends with some conclusions.

\section{Related Work}
Comparing to traditional video which encodes a view from a single perspective, 360$^{\circ}$ videos encode a full, omnidirectional view. Streaming the entire view of 360$^{\circ}$ videos leads to a considerable waste of the network bandwidth, as users utilize only a small portion, which contains their viewport, of the full image. Researchers have proposed some notable algorithms to reduce the bandwidth requirement.


A promising idea is to spatially split 360$^{\circ}$ videos into tiles, and encode each tile in multiple bitrates. Users can adaptively fetch tiles of different quality according to their network conditions.
A few research, like \cite{8576614}, choose to explore the cooperate chance between several VR users that can minimize the bandwidth consumption.
Most still choose to dedicate to single user streaming. Extending MPEG-DASH SRD to the 3D space, Mohammad Hosseini\cite{Hosseini2017View} showcases a dynamic view-aware adaptation technique to tackle the high bandwidth demands of streaming 360$^{\circ}$ videos to wireless VR headsets. Chao Zhou et al.\cite{8486282} select tiles by solving a set of integer linear programs independently on clusters of collected user views. 
Yu Guan et al.\cite{guan2019pano} build a new quality model that leverages three 360$^{\circ}$ video-specific factors. Based on the model a variable-sized tiling scheme is proposed to strike a balance between the perceived quality and video encoding efficiency. In \cite{jiang2018plato}, Xiaolan Jiang uses RL to take advantage of historical viewport and bandwidth information, to make optimal adaptive decisions of tile quality.

\section{System Design}
In this section, we firstly present the system architecture of OFB-VR. Secondly, we introduce each part in detail, including the optical flow based $360^\circ$ specialized QoE metric PSNR-OF, the versatile-size tiling scheme that can cooperate with it, and the architecture of the ABR RL network.
\subsection{System Model}
The workflow of OFB-VR is depicted in figure \ref{Flowchart of OFB-VR}. The system can be divided into two parts, the edge server part and the client part. At the edge server part, firstly, original videos uploaded from capture devices to the edge computation server are cut into chunks, and each chunk is split into basic tiles. Secondly, we introduce a CNN-based optical flow estimation  network to extract pixel-wise motions. Then, the PSNR-OF calculation module takes users' viewpoint motion, each tile's size in each quality level, and optical flow estimation result as input, to compute each basic tile's PSNR-OF efficiency. Finally, according to the PSNR-OF efficiency scores, adjacent basic tiles with similar scores are merged and encoded into the final version to be stored in the edge content distribution server. In the client part, the ABR RL network firstly acquires the viewpoint of the user and computes the corresponding PSNR-OF scores of each tile. Then make the optimal tiles' quality choice according to its trained policy, aiming at maximizing the weighted sum of target parameters. Detailed descriptions of each part can be found in the following subsections. 

\subsection{Optical Flow Estimation Based 360$^{\circ}$ QoE Awared Quality Metric}
\subsubsection{Optical Flow Based Quantification of Users' Awareness of Quality Distortion}

When watching a $360^\circ$ video, the user's attention can be attracted by a moving target and move rapidly to follow it. Thanks to this unique and crucial characteristic, background distortion in a $360^\circ$ video can escape from the user's awareness under some specific condition. In this paper, we take relative viewpoint velocity, and depth difference between the attached object and background as our quantify factor, to seize chances that can transfer parts of the video in low quality without being noticed.

\textbf{Relative viewpoint velocity:} It's a simple fact that humans can't perceive a moving target as clear and distinct as a static one. When watching regular videos, it's impossible to attain the user's viewpoint information. As a result, the encoder can merely assume that users' attention distributed evenly and encode each pixel equally. However, VR has the inherent advantage of knowing users' viewpoint at any time. Lowering the quality of parts, whose motion is dramatically different from the user's viewpoint motion, is a feasible and promising way to save bandwidth consumption without noticeable quality decrease. For example, if the user's attention is strictly attached to a moving object, the object would seem stationary, yet the background would seem moving backward. Under that circumstance, blur and compression in the relative moving part are hard to notice.
    
    
\textbf{Relative depth difference:} Similar to the relative viewpoint, depth differences between the focused area and other parts of the video can also cripple humans' sensitivity about quality decline. For example, if the user pays more attention to a close object, namely the foreground object, background elements can maintain QoE while decreasing bitrate.

\textbf{Quantification of Users' Quality Loss Awareness:}
Above mentioned two factors are subjective and ambiguous. With the notion and empirical fitting function of JND, researchers can quantify the impacts of those factors. However, it remains challenging to precisely extract depth information and pixel-wise motion for its inherent complexity in the algorithmic aspect. In this paper, we introduce real-time optical flow estimation in JND estimation, named JND-OFE.

JND-OFE is defined as the minimal changes in pixel values that can be noticed by viewers. JND-OFE formula can quantify the threshold of quality drop that users can't notice. In this paper, we carry over the empirical fitting function from \cite{zhao2011video}, then unify the unit and range of our parameters to assure its validation.

JND-OFE of Relative viewpoint velocity, named SJND:
\begin{equation}
\operatorname{SJND_{i,j}}(\Delta v)=2.047\times \Delta v^{0.634}+8
\label{SJND}
\end{equation}

JND-OFE of Relative depth difference, named DJND:
\begin{equation}
\operatorname{DJND_{i,j}}(\Delta d) = \left\{
\begin{array}{rcl}
9\times \Delta d + 12 & & {\Delta d < 1}\\
29\times \Delta d - 8 & & {\Delta d \geq 1}
\end{array} \right.
\end{equation}
Inspired by Pano\cite{guan2019pano}, we take the impact of these two factors of JND as independent. So the joint JND-OFE can be expressed as: $\operatorname{JND-OFE_{i,j}}(\Delta v, \Delta d)= \operatorname{SJND_{i,j}}(\Delta v) \times \operatorname{DJND_{i,j}}(\Delta d)$


To leverage the panoramic quality-determining factors, we design a motion and depth extract module based on optical flow estimation. With the optical flow map $\vec{F}$ and velocity of viewpoint $\vec{v}$, all it takes is to calculate$\begin{vmatrix}\vec{F}_{i,j} - \vec{v}\end{vmatrix}$ to get the relative viewpoint velocity map $\Delta V$. As for depth estimation, it's a little complicated. Although the velocity map isn't equivalent to the depth map, it's possible to distinguish background elements with foreground elements. Videos have at least 24 frames per second(FPS), which guarantees that the time gap between two consequent frames can't be longer than 41.6ms. In such a short period, objects at distant, no matter moving or static, seem static. Based on this, according to $\vec{F}$, a mask of background can be concluded. Then, normalize all elements and set the value of background elements to 0 to get velocity depth base $D$. After that, take the difference map $\Delta D$ between $D$ and $\vec{v}$ as the output of relative velocity depth. The last step is to respectively calculate SJND and DJND and produce the joint JND result.
\subsubsection{PSNR-OF, A QoE Metric concerning users' awareness of quality distortion}
OFB-VR proposes a new QoE metric specifies on $360^\circ$ videos. The new metric is based on Peak Signal-to-Noise Ratio (PSNR) and the proposed quality loss awareness factors. With real-time optical flow estimation, we can accurately acquire users' tolerance upper bound of quality distortion, the JND-OFE score. We insert JND-OFE in PSNR's MSE calculation. When calculating the difference between an original pixel and its encoded version, results lower than the JND-OFE threshold are ignorable. The new metric is referred to as Peak Signal-to-Noise Ratio through Optical Flow (PSNR-OF), which can ignore the impact of quality distortions in different moving direction or depth layer. In other words, PSNR-OF can evaluate panoramic videos in a way that accord to intuitive human feelings. PSNR-OF can be presented as follow:
\begin{equation}
    \operatorname{PSNR-OF}(l) = 20 \times \log_{10}\frac{2^{N}-1}{\sqrt{\operatorname{MSE}(l)}}
\end{equation}
\begin{equation}
    \operatorname{MSE}(l) = \frac{\sum_{i}^{row}\sum_{j}^{col}
    \begin{bmatrix}(
        \begin{vmatrix}
            I_{i,j}(l)-\hat{I}_{i,j}(l)
        \end{vmatrix}) \times \operatorname{k}_{i,j}(l)
    \end{bmatrix}^2}{col \times row}
\end{equation}
\begin{equation}
    \operatorname{k}_{i,j}(l) = \frac{\operatorname{sgn}(
        \begin{vmatrix}
            I_{i,j}(l)-\hat{I}_{i,j}(l)
        \end{vmatrix} - \operatorname{JND-OFE}_{i,j})+1}
        {2}
\end{equation}
Generally, for uint8 data, N = 8. In the formula. m and n respectively denote rows and columns of the original image. $I_{i,j}$ stand for the pixel value at $(i,j)$ of the original image, and $\hat{I}_{i,j}(l)$ for that of image encoded at quality level $l$. 

\subsection{Versatile-size Tiling Scheme}
After calculating the pixel-wise PSNR-OF, it's sufficient to carry on to perform the versatile-size tiling scheme. Firstly, fine-grained split the original chunk into basic tiles with a $12\times24$ grid and compute mean PSNR-OF in each tile as tile-wise PSNR-OF. Secondly, compute PSNR-OF efficiency $E$ of each tile, which stands for average PSNR-OF increase per quality level. PSNR-OF efficiency can be denoted as:
\begin{equation}
    E_{i,j} = \frac{\operatorname{PSNR-OF_{i,j}}(l_{high})-\operatorname{PSNR-OF_{i,j}}(l_{low})}
        {l_{high}-l_{low}}
\end{equation}
Finally, with the tile-wise efficiency $E$, we can group similar tiles. The overall goal is to group 288 basic tiles into fixed number K, say 50, of coarse-grained tiles while minimizing PSNR-OF efficiency's total variance. In general, our algorithm uses the rectangle of the whole frame as a beginning. It iteratively chooses the best cut among all possibilities, horizontally or vertically. It repeats the same procedure to cut two tiles until there are K tiles. The detail implementation is described in algorithm 1. 

\begin{algorithm}[htb]
    \caption{Versatile-size Tiling Algorithm}  
    \begin{algorithmic}
        \State \Call{$Tiling$}{$1,12,1,24$}
        \State
        \Function{$Tiling$}{$x_1, x_2, y_1, y_2$}
            \State$cutDirection, cutPosition, availableSteps\gets$\\    \Call{$GetBestCut$}{$x_1, x_2, y_1, y_2$}
            \If{$availableSteps == 0$}
                \State \Return
            \EndIf
            \If{$cutPosition == horizontal$}
                \State \Call{$Tiling$}{$x_1, cutPosition, y_1, y_2$}
                \State \Call{$Tiling$}{$cutPosition+1, x_2, y_1, y_2$}
            \Else
                \State \Call{$Tiling$}{$x_1, x_2, y_1, cutPosition$}
                \State \Call{$Tiling$}{$x_1, x_2, cutPosition+1, y_2$}
            \EndIf
        \EndFunction
        \State
        \Function{$GetBestCut$}{$x_1, x_2, y_1, y_2$}
            \State Calculate best $cutPosition$ and $cutDirection$ that can minimize PSNR-OF variance;
            \State Renew $availableSteps$;
            \State \Return{$cutDirection, cutPosition, availableSteps$}
        \EndFunction
    \end{algorithmic}  
\end{algorithm}  

\subsection{Reinforce Learning ABR network}
In this paper, we implement the state-of-the-art actor-critic method, A3C, as our training algorithm. The A3C algorithm uses two networks, an actor and a critic. The actor takes action $\pi_\theta(s)$ according to current state $s$ to maximize the cumulative reward. The critic evaluates actions from the actor and helps it to update its policy network parameters $\theta$.

Training a network to decide the tile-wise quality level is tough because of the enormous action space. To be more specific, if each tile has 6 possible versions of different quality, and the video is split into 12 columns and 6 rows, namely 72 tiles, the size of its action space would be $5^{72}$. So we propose a simplified alternative, all tiles are sorted into three types: Areas that near the viewpoint, which are referred to as core areas; Areas that surround the core area, which are referred to as surrounding areas; Areas that users can't see, which are referred to as outside areas. Our network in charge of giving the overall bandwidth upper bound of each area. Tile-wise quality selection tasks are left to the local dynamic programming (DP) algorithm. Accordingly, our state $s_t$, namely network input, is defined as:
\begin{equation}
    s_t = \left( \vec{P_t},\vec{\tau_t},\vec{co_t},\vec{su_t},\vec{out_t}, buf_t, ratio_t, acc_{out}\right)
\end{equation}
Where $\vec{P_t}, \vec{\tau_t}, \vec{co_t}, \vec{su_t}, \vec{out_t}$ respectively stand for PSNR-F, chunk download time, core area total bitrate, surround area total bitrate, outside area total bitrate. Each of them is a $m$ dimensional vector that accordingly records its historical data. $m$ intends the referred historical data's period length. $buf_t$ is the available buffer length, $rate_t$ is the actual mean bitrate while downloading the last chunk, $ratio_t$ is the ratio of basic tiles that predicted to be outside but present in the real viewport. In our network, there are 5 identical 1D-CNN with 128 filters to receive inputs, each of size 3 with stride 1, and other 3 full connected NN with 128 neurons to receive $buf_t, rate_t, acc_{out}$. A full-connected layer with N neurons is responded to receiving outputs from these layers, where N is the size of action space, namely the overall possible quantity of QP choice of each tile. The critic network shares a similar NN structure with the actor network except for the output layer. The critic network only has one neuron as its output.

We define a total reward function to reflect the value of current action, concerning above mentioned PSNR-OF as our main reward factor. Rebuffer ratio, a typical video streaming QoE metric as our main penalty factor. Besides, we take outside area ratio into account as a penalty factor as well, since the outside area is meant to be transferred in relatively low quality or even not to be used at all. The reward function can be denoted as:
\begin{equation}
\begin{aligned}
    reward = &\sum_{n = 1}^{N}P_n -  \alpha \sum_{n = 1}^{N}Rt_n - \beta\sum_{n = 1}^{N}ratio_n \\
    & - \sum_{n = 1}^{N-1}\left | P_{n+1} - P_n \right |
\end{aligned}
\end{equation}

For a video with N chunks, $P_n$ is the unified mean PSNR-OF score of the n rd chunck. $Rt_n$ stands for the rebuffering time of the n rd chunk while downloading. $ratio_n$ is the outside area ratio of the n rd chunk. We use this reward function as the metric to evaluate the actor network's decision.

A3C is a policy gradient training method. It uses the gradient to help update policy parameters. Given the policy network parameters $\theta$, the gradient can be denoted as:
\begin{equation}
    \nabla_\theta R(\theta) = \mathbb{E}_{\pi_\theta} \left[ \nabla_\theta log\pi_\theta \left(s,a\right)A^{\pi_\theta} \left(s,a\right)\right]
\end{equation}
$A^{\pi_\theta}(s,t)$ is the advantage function:
\begin{equation}
    A^{\pi_\theta}(s,t) = r_t + \gamma r_{t+1}+...+\gamma^{n-1} r_{t+n-1}+\gamma^n V^{\pi_\theta}(s') -V^{\pi_\theta}(s) 
\end{equation}
where $V^{\pi_\theta}(s)$ is the output of the critic network in state s and policy parameters $\theta$. $s'$ is the new state after taking action $\pi_\theta(s)$. With the gradient function, the actor network's parameters $\theta$ can be updated with learning rate $\alpha$:
\begin{equation}
    \theta = \theta + \alpha\nabla_\theta log\pi_\theta(s,a)A^{\pi_\theta}(s,t)
\end{equation}
\section{Experiemts and Evaluation}
In this section, we firstly provide set up of our videos, viewpoint, and bandwidth datasets, and parameters of our RL network. Then we evaluate OFB-VR's performance with both user study and viewpoint-based simulation. User study aims to examine whether PSNR-OF can fit intuitive human feelings and form a trace database for followed up simulations. After that, we perform a viewpoint-based simulation in ideal and stable bandwidth conditions to quantitatively evaluate the improvement of our method on bandwidth saving and QoE enhancing. Finally, we train our RL network with practical bandwidth datasets to evaluate the overall performance and compare it with baseline schemes.
\subsection{Experiment Setup}
\textbf{video and viewpoint dataset: }Our video database consists of 20 360° videos that add up to over 60 minutes. Each video's resolution is 2880*1440, and FPS is 30. 

Each videos is encoded into 5 quality levels, QP = 20, 25, 30, 35, 40, of one-second chunks with the x264 encoder for subsequent procedures. Our 10 volunteers are asked to watch those videos and record their viewpoint trace respectively, as our viewpoint datasets.

\textbf{ABR Reinforce Learning network parameters and dataset: }In the state input part, we set dimension of input vectors $m = 8$, to recall past 8 seconds' experience. Possible quality level choices number to 6, including qp(20, 25, 30, 35, 40) and a blank level. As a result, the number of action space is $6^3$, since each area has 6 possible choices.In the reward function, we set $\alpha = 33$, $\beta = 1$.During training, we set the discount factor $\gamma = 0.99$. The learning rates for the actor and the critic are both set to 0.0001.

We use a dataset from\cite{7546928}. The dataset contains available bandwidth in real 4G/LTE networks within the city of Ghent, Belgium, from 2015-12-16 to 2016-02-04. It includes bandwidth tracks recorded from different transportation, car, train, tram, bus, bicycle and foot. We adopt this dataset mainly for its drastic variation in bandwidth, which can test the robustness of our trained network. The throughput grows up to 95Mbps at some cases, and drop to severed limited when occasionally switching to 3G. However, the mean throughput of the dataset is up to 35Mbps, which is so adequate for single video streaming that can even streaming videos from our datasets in full size. So we scale down the dataset to two lower mean throughput version. The higher one has a mean throughput of 5Mbps, and the lower one has a mean throughput of 2Mbps. There are 40 logs in total, and each has a relatively limited duratio ranging from 166 to 758 sec. During traning we iterately chose a random track in the bandwidth dataset to overcome the limitation.

\textbf{Baseline setup}
We take two recently proposed DASH based VR streaming scheme, Plato and Pano, as baselines. Both of them are viewport-driven as well. Pano is a variable-size tiling scheme with traditional JND and DP ABR. We reconstruct its DP ABR part, to make sure it's compatible with the proposed RL ABR system. Plato is a typical fixed-size tiling scheme. Considering that it doesn't make use of the user's quality awareness, to make it comparable, we reconstruct it with the PSNR-OF metric. However, despite the introduction of JND-OFE, its JND-OFE factor is always 1, according to equation 5. So, in fact, PSNR-OF in Plato degrade to regular PSNR.

\subsection{Validation of Optical Flow Based JND}
\begin{figure}[h]
\begin{minipage}[t]{0.49\linewidth}
\includegraphics[width=1\textwidth]{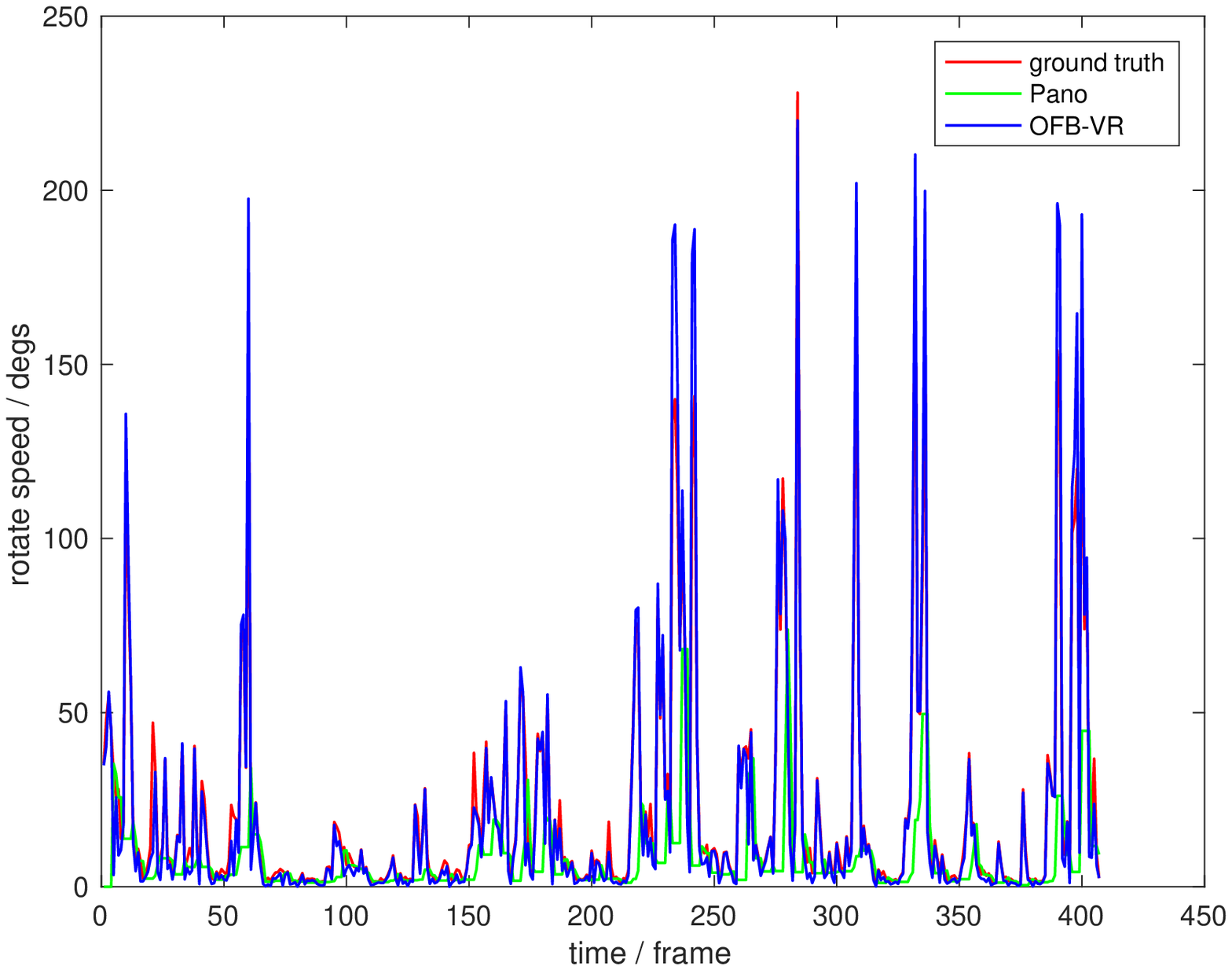}
\caption{Speed Estimation Error \label{speed_error}}
\end{minipage}
\hfill
\begin{minipage}[t]{0.5\linewidth}
\centering
\includegraphics[width=1\textwidth]{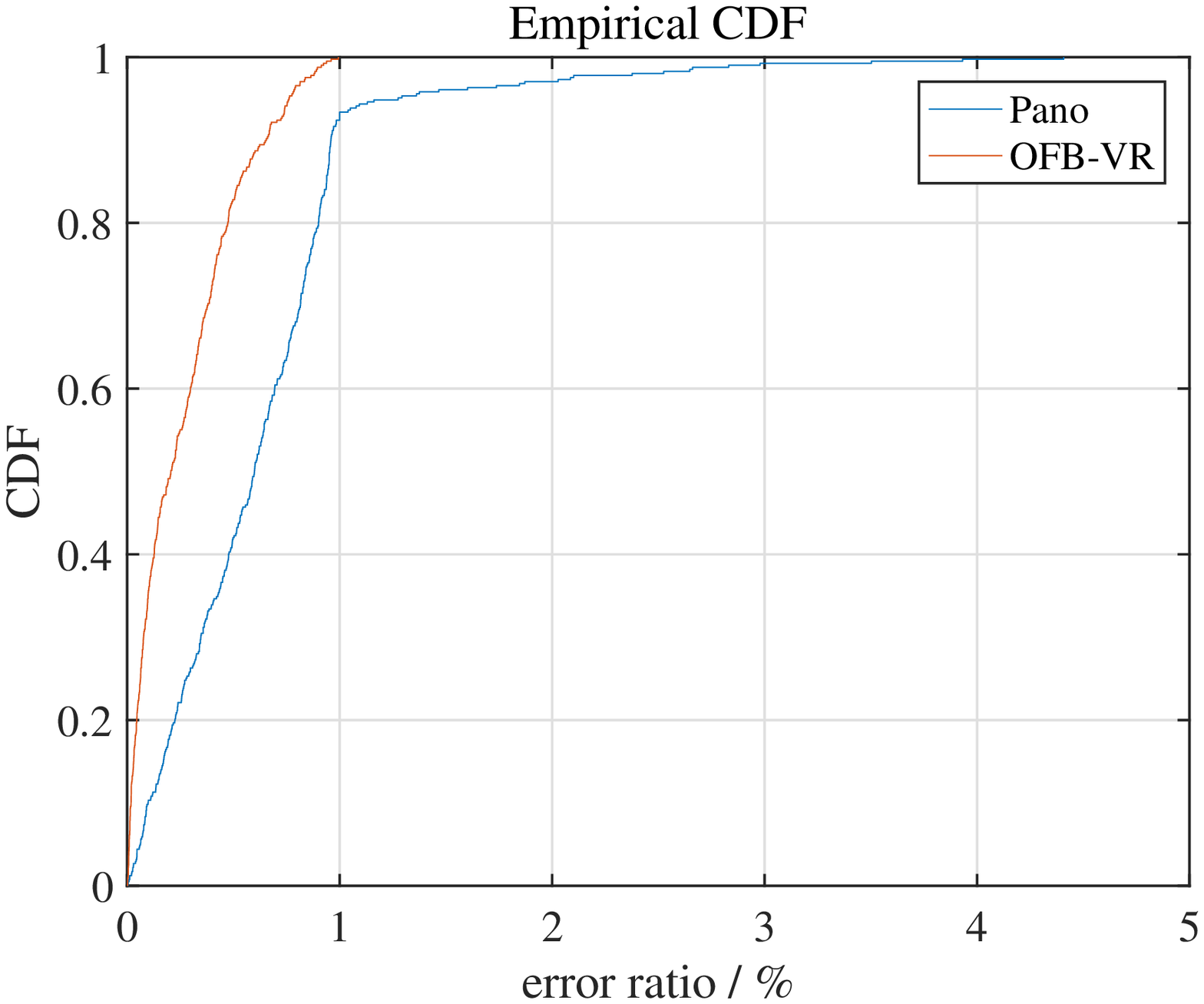}
\caption{CDF of Speed Estimation Error\label{speed_error_cdf}}
\end{minipage}
\end{figure}
In OFB-VR, the accuracy of JND estimation strictly depends on the accuracy of motion field estimation. As a result, it is sufficient to evaluate JND estimation results by analyzing the result of the speed estimation result. Specifically, in our scheme, it is relative speed derived from optical flow estimation, while in traditional JND, like Pano, it is derived from CNN-based object detection. Because CNN-based object detection's inherent limitation in generalization, it can misjudge or even can't recognize some untrained target at all. As a result, Pano chooses to use the relative speed lower bound in recent history as a rough substitute of a precise pixel-wise relative viewpoint. As for Plato, to make it comparable, we unify the QoE metric with PSNR-OF. However, since it doesn't introduce any QoE concerned notion, its JND-OFE value is always set to 0, namely, it is absent in the comparison of relative value accuracy.  

In figure \ref{speed_error}, though optical flow estimation error can lead to variant errors of relative speed calculation, our method still outperforms the conservative prediction made by speed lower bound. The improvement is evident in the speed estimation error's CDF, figure \ref{speed_error_cdf}. No doubt, the more accurate the JND estimation is, the better the performance in bandwidth saving or QoE it can get.
\subsection{Improvement on Bandwidth Consumption and QoE}
    
\begin{figure}[h]
\begin{minipage}[t]{0.5\linewidth}
\includegraphics[width=1\textwidth]{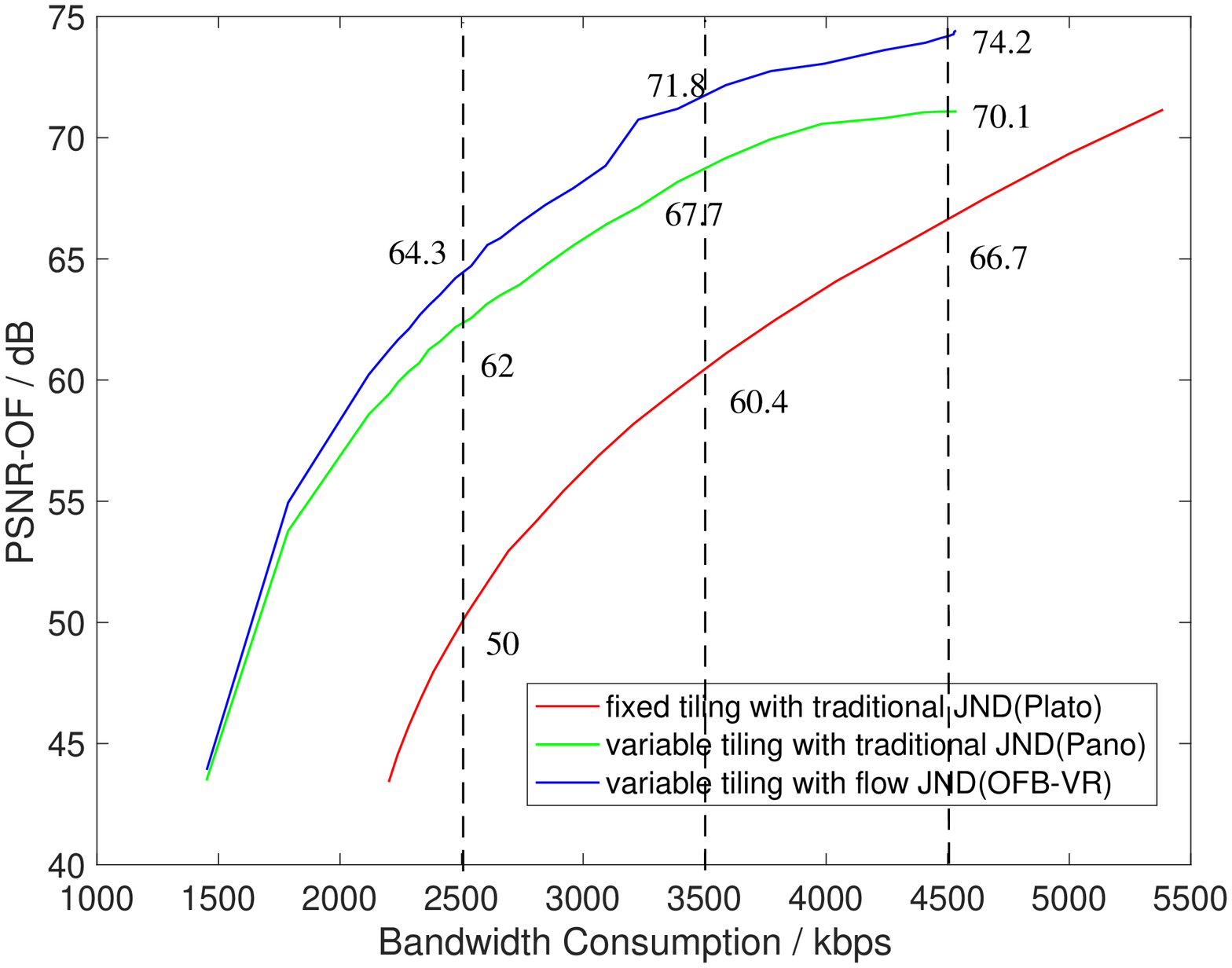}
\caption{Overall QoE Improvement\label{overall_PSNR}}
\end{minipage}
\hfill
\begin{minipage}[t]{0.49\linewidth}
\centering
\includegraphics[width=1\textwidth]{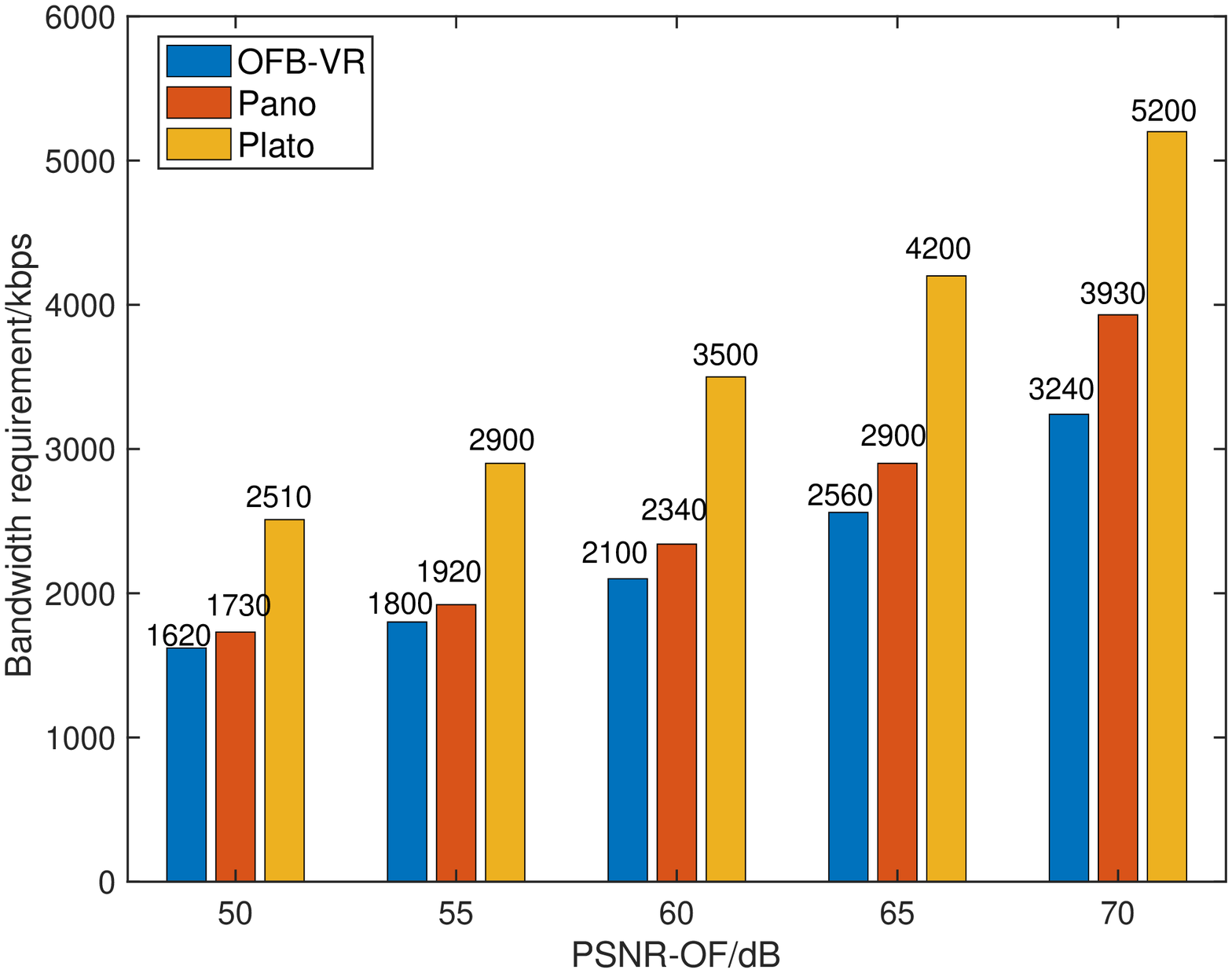}
\caption{Overall Bandwidth Saving\label{overall_MOS}}
\end{minipage}
\end{figure}

In this part, we make an simulation on three VR streaming scheme with stable bandwidth, including OFB-VR, Pano, and Plato. We run this simulation in order to analysis theoretical joint improvement of proposed versatile-size scheme and PSNR-OF. To ensure fair compression, We unify the adaptive logic and viewpoint prediction method. The result is concluded in the same video and trace database.

As depicted in figure \ref{overall_PSNR}, ORB-VR can increase PSNR-OF at most by 4 dB in comparison with Pano, or 7.4 dB in comparison with Plato, while no extra bandwidth is required. It's evident that the fixed-tiling scheme severely harms Plato's performance.it is competitive only when the bandwidth is adequate to stream each tile in high quality. According to figure \ref{overall_MOS}, to maintain fundamental PSNR-OF score, OFB-VR and Pano need a similar bandwidth, because JND-OFE can only make a constrained contribution with limited bandwidth. However, to present the user with medium or high quality, say 70 PSNR-OF scores, OFB-VR can save up to 17.6\% bandwidth in comparison with Pano. while comparing with Plato, to achieve the same PSNR-OF score, OFB-VR only needs about 60\% bandwidth.
\subsection{Performance of the ABR RL network}
\begin{figure}[h]
\begin{minipage}[t]{0.5\linewidth}
\includegraphics[width=1\textwidth]{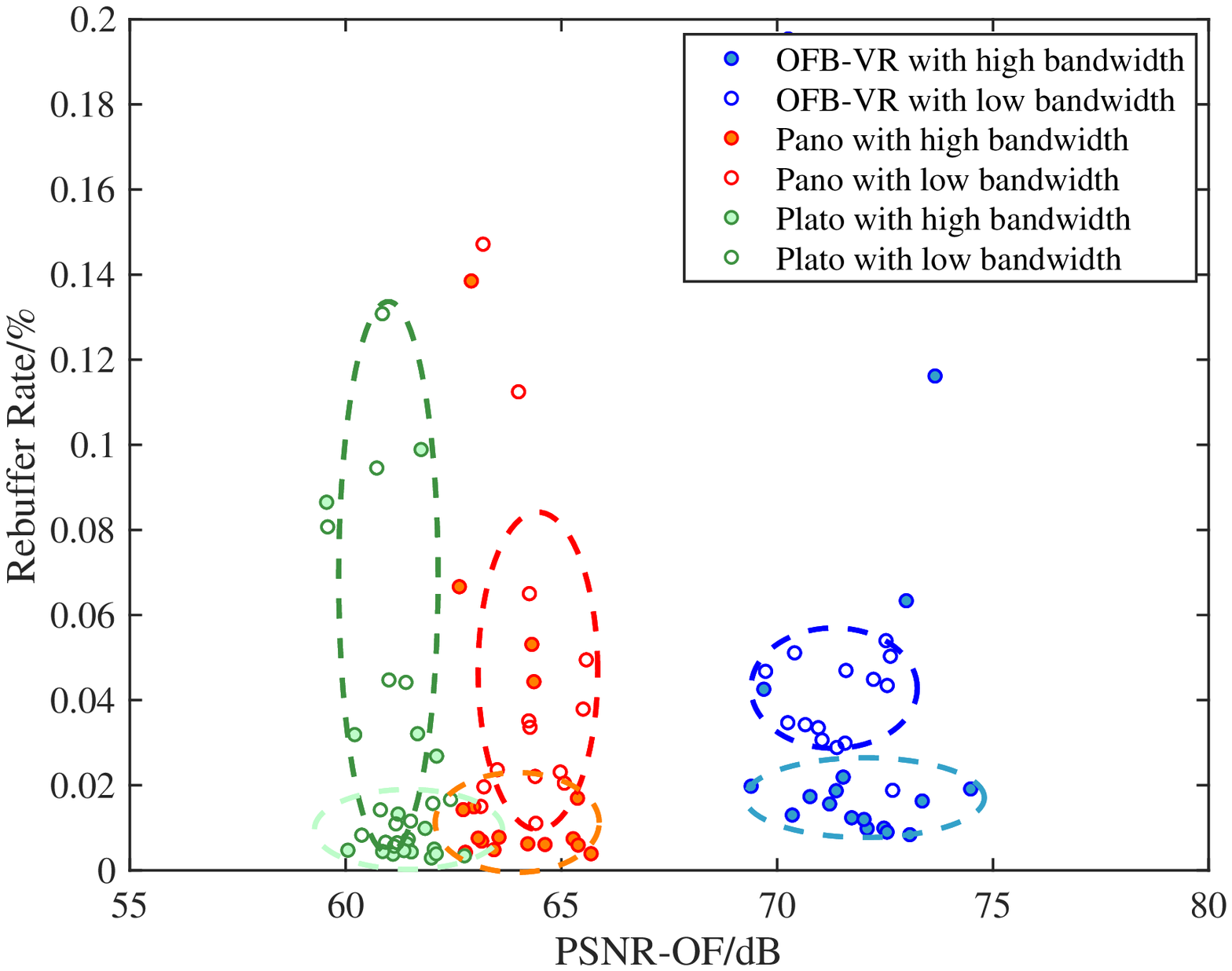}
\caption{ABR Reinforce learning network performance\label{RL_performance}}
\end{minipage}
\hfill
\begin{minipage}[t]{0.49\linewidth}
\centering
\includegraphics[width=1\textwidth]{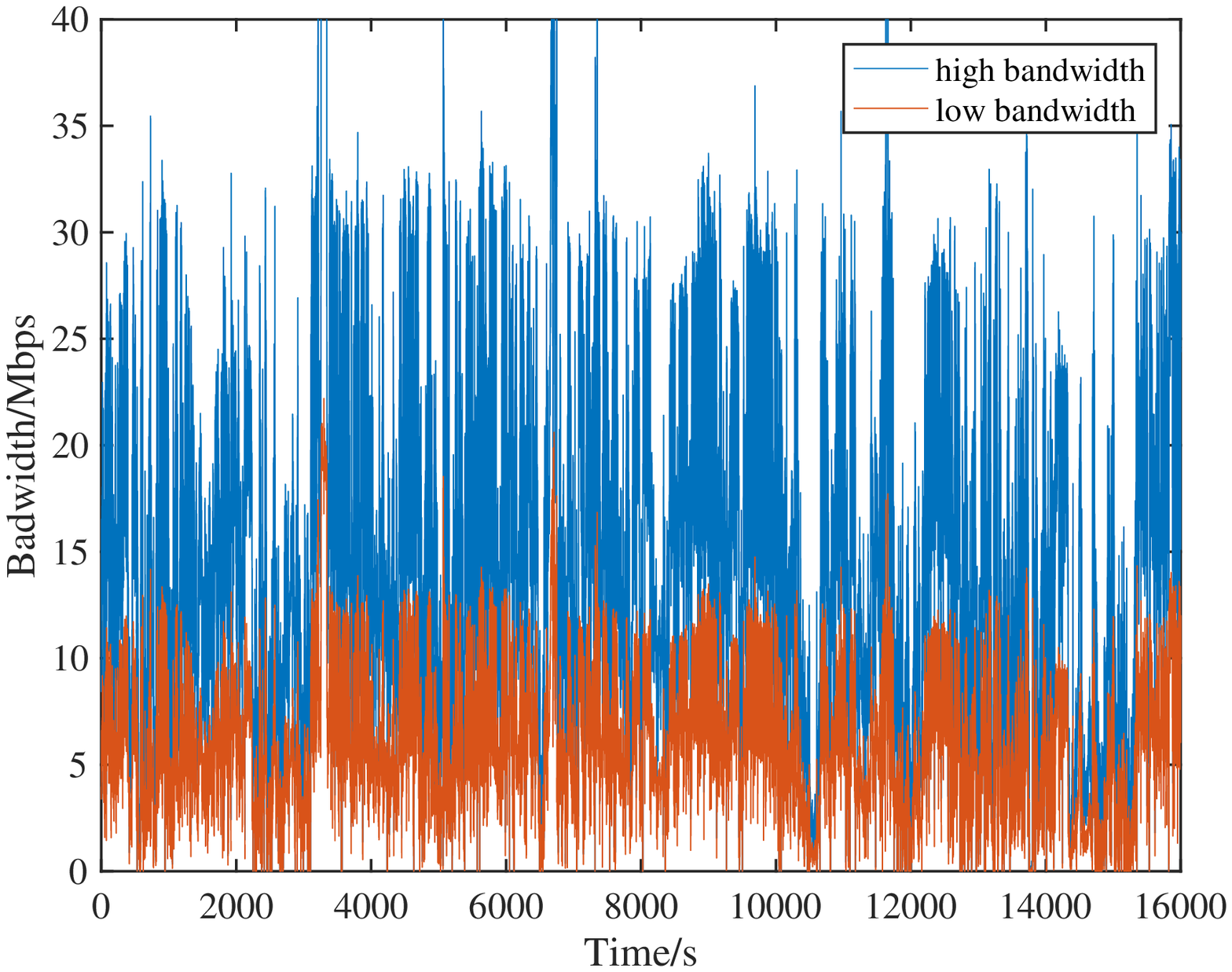}
\caption{Sample of bandwidth datasets\label{bandwidth}}
\end{minipage}
\end{figure}
After the independent test of the QoE evaluation part, we integrate them into the proposed ABR RL module, to test the overall performance of the proposed system. In this part, since we use the practical LTE bandwidth dataset to take the place of ideal and fixed bandwidth, we jointly take the rebuffer rate and PSNR-OF as our evaluate metric. To carry on the test, for each scheme, we randomly feed one video's PSNR-OF and viewpoint from our video dataset to the RL network. Then iteratively repeat the procedure until all video is tested.

Figure \ref{bandwidth} is the practical bandwidth track we used. The two tracks are derived from the original dataset with different scale down proportions. The lower one only has 5Mbps drastically changed bandwidth, while the higher one has an adequate bandwidth of 10Mbps, but it fluctuates significantly as well.

The performances of the ABR RL network with three difference schemes are respectively given in figure \ref{RL_performance}. The figure presents the PSNR-OF score and rebuffer rate jointly. Each point reflects the mean PSNR-OF and mean rebuffer rate during one video's processing. The closer the distance between the right-bottom and the point, the better the general performance the scheme has. Mean PSNR-OF scores of three schemes are 71.09, 64.90, and 61.4, respectively. In other words, our system can increase the mean PSNR-OF score by 9.5-15.8\% while maintaining the same rebuffer ratio compared with Pano and Plato in a fluctuate LTE bandwidth dataset. It is obvious that the proposed OFB-VR outperforms Pano in PSNR-OF while maintaining the rebuffer ratio at the same level.

\section{Conclusion}
By leveraging relative velocity and depth difference as QoE awareness quantify factors, OFB-VR extends the traditional video QoE metric, PSNR, to seize the user's awareness of QoE loss. With the help of PSNR-OF, the versatile-size tiling scheme is proposed to take the place of the fixed-size tiling scheme by grouping tiles with similar PSNR-OF efficiency together. Base on this, an ABR RL network is designed to adaptively make optimal choices of each tile's quality for streaming. The performance of the proposed system is evaluated on a real LTE bandwidth database. The experiment result proves the validation of the proposed optical-flow-based JND model, and quantitively evaluate the improvement of our method over two baselines. It turns out that our method is useful to save bandwidth resources, especially when providing users with high-quality panoramic content.

\bibliographystyle{IEEEtran}

\end{document}